\documentclass[a4paper,11pt]{article}
\usepackage{pos}
\usepackage{booktabs}
\usepackage{siunitx}
\usepackage[capitalize]{cleveref}

\DeclareSIUnit\barn{b}
\newcommand{\rmd}{\mathrm{d}}
\newcommand{\LO}{\ensuremath{\mathrm{LO}}}
\newcommand{\NLO}{\ensuremath{\mathrm{NLO}}}
\newcommand{\NNLO}{\ensuremath{\mathrm{NNLO}}}
\newcommand{\NNNLO}{\ensuremath{\mathrm{N}^3\mathrm{LO}}}
\newcommand{\LOPS}{\ensuremath{\mathrm{LO}{+}\mathrm{PS}}}
\newcommand{\MiNNLOPS}{\ensuremath{\mathrm{MiNNLO}{+}\mathrm{PS}}}

\title{Parton-shower and fixed-order QCD effects in Higgs boson production in 
  weak-boson fusion and its decays to bottom quarks}
\ShortTitle{Higgs production in weak-boson fusion and $H \to b\bar{b}$ decays}

\author*[a]{Arnd Behring}

\affiliation[a]{Max-Planck-Institut für Physik \\
  Boltzmannstraße 8, 85748 Garching, Germany}

\emailAdd{abehring@mpp.mpg.de}

\abstract{We study the impact of higher-order QCD corrections on Higgs
  production via weak-boson fusion with subsequent decays of the Higgs boson to
  bottom-quark pairs. We consider realistic fiducial cuts and observe large
  corrections in a fixed-order calculation. They arise from the interplay of
  the particular fiducial cuts considered and the radiative corrections to the
  Higgs decay subprocess. Switching to a parton-shower-matched description of
  the decay captures a significant portion of the corrections already at the
  lowest order and improves the perturbative convergence of the calculation.}

\FullConference{The 33rd International Workshop on Deep Inelastic Scattering and Related Subjects (DIS2026)\\
4 - 8 May 2026\\
Bologna, Italy\\}

\begin{document}
\maketitle

\section{Introduction}
While Higgs boson production via weak-boson fusion (WBF) is only the second
largest production channel for Higgs bosons at the Large Hadron Collider, it is
nevertheless an important avenue to precisely measure properties of Higgs
bosons.
In contrast to the most prevalent production channel, gluon fusion, the final
state of the WBF process contains two hard forward jets besides the Higgs boson
($p p \to H j j$), which allows for a separation of the signal process from
other backgrounds.
This is particularly important when considering decays of the Higgs boson to
bottom-quark pairs ($H \to b \bar{b}$), since direct QCD production of $b$
quarks without going through an intermediate Higgs boson would very much
overwhelm the signal if not for the additional forwards jets.
In turn, Higgs boson decays to bottom quarks can be used, e.g., to measure the
Yukawa coupling of Higgs bosons to bottom quarks.

Both the production and decay processes have been calculated very precisely.
The inclusive cross section for WBF Higgs production is known up to
next-to-next-to-next-to-leading order ($\NNNLO$) in QCD~\cite{Dreyer:2016oyx}
using the factorised (structure function) approximation, while fully
differential calculations in the same approximation are available up to
next-to-next-to-leading order ($\NNLO$)~\cite{Cacciari:2015jma,%
Cruz-Martinez:2018rod,Asteriadis:2021gpd}. Moreover, electroweak corrections
and interference effects are known to next-to-leading order
($\NLO$)~\cite{Ciccolini:2007jr}. Non-factorisable corrections, involving
gluon exchanges between the fermion lines, are known at $\NNLO$ in
QCD~\cite{Liu:2019tuy,Dreyer:2020urf,Asteriadis:2023nyl}.
Fully differential predictions for the decay of the Higgs boson exist for
massless $b$ quarks up to $\NNNLO$ in QCD \cite{Mondini:2019gid} and for
massive quarks up to $\NNLO$ in
QCD~\cite{Bernreuther:2018ynm,Behring:2019oci}.

Here, we summarise the findings of Ref.~\cite{Asteriadis:2024nbg}, where we
combined the $\NNLO$ QCD description for WBF Higgs production with an $\NNLO$
QCD fixed-order description for Higgs decays to bottom-quark pairs, as well as
a follow-up study~\cite{Behring:2025msh}, in which we upgraded the description
of the Higgs decays to an $\NNLO$ calculation matched to parton showers.

\section{Setup}
\label{sec:setup}
In order to separate signal from background, experimental analyses apply fairly
stringent kinematic cuts on both the forward jets from the production process
and on the $b$ jets produced in the decay of the Higgs boson.
For our calculation, we emulate an analysis by the ATLAS
experiment~\cite{ATLAS:2020bhl}. All jets are defined using the anti-$k_T$
algorithm using a jet radius parameter of $R=0.4$.
We demand two WBF-tagging jets, i.e., light, non-$b$-flavoured jets with the
largest and second-largest transverse momenta and rapidities $|y_j| < 4.5$.
They must individually have a transverse momentum of at least
$p_{\perp,j} > \qty{25}{\GeV}$ and together an invariant mass of
$m_{j_1 j_2} > \qty{600}{\GeV}$. The two jets have to have a rapidity separation
of at least $|y_1 - y_2| > 4.5$. Combined with the rapidity requirement on the
individual jets, this implies that the two jets lie in opposite hemispheres.
In addition, we require a pair of $b$-jets in the central region, i.e., with
rapidities $|y_b| < 2.5$, and a transverse momentum of at least
$p_{\perp,b} > \qty{65}{\GeV}$. If there is more than one candidate for such a
pair of $b$-jets, the one whose invariant mass is closest to $m_H$, the mass of
the Higgs boson, is chosen.
Note that the cut on the transverse momentum of the $b$ jets implies that Higgs
bosons decaying at rest would not pass this cut, as $p_{\perp,b}^\text{cut} >
\frac{m_H}{2}$. We will come back to this observation below.

We work in the narrow-width approximation which allows us to write the
differential cross section as
\begin{align}
  \rmd\sigma
    &= \rmd\sigma_\text{WBF} \,
       \mathrm{Br}_{H \to b\bar{b}} \,
       \frac{\rmd\Gamma_{H \to b\bar{b}}}{\Gamma_{H \to b\bar{b}}}
  \,, &
  \text{where}~
  \mathrm{Br}_{H \to b\bar{b}}
    &= \frac{\Gamma_{H \to b\bar{b}}}{\Gamma_H}
  \,.
\end{align}
Here, $\rmd\sigma_\text{WBF}$ denotes the differential cross section of the
production subprocess $p p \to H j j$ and $\rmd \Gamma_{H \to b\bar{b}}$ stands
for the differential decay width of the Higgs boson to $b$ quark pairs.  The
inclusive partial decay width for $H \to b\bar{b}$ is given by $\Gamma_{H \to
b\bar{b}}$ and the total decay width of the Higgs boson is denoted by
$\Gamma_H$.
This conveniently separates the description of the production and decay
subprocesses.

\section{Fixed-order calculation}
\begin{table}
  \centering
  \begin{tabular}{ccccc}
    \toprule
    $\sigma/\unit{\femto\barn}$ & fixed order~\cite{Asteriadis:2024nbg} & $\LOPS$ \cite{Behring:2025msh} & $\MiNNLOPS$ \cite{Behring:2025msh} \\
    \midrule
    $\LO$                       & $75.6\hphantom{(1)}$                  & $46.6\hphantom{(1)}$           & $45.2\hphantom{(1)}$               \\
    $\NLO$                      & $52.4\hphantom{(1)}$                  & $43.6(1)$                      & $42.3\hphantom{(1)}$               \\
    $\NNLO$                     & $44.6(1)$                             & $43.1(1)$                      & $41.4(1)$                          \\
    \bottomrule
  \end{tabular}
  \caption{
    Fiducial cross sections subject to the cuts outlined in \cref{sec:setup}
    for WBF Higgs production with subsequent decays to $b$-quark pairs.
    The results are given for different approximations for the production
    (rows) and for the decay (columns) subprocesses.
    The error shown corresponds to the Monte-Carlo integration uncertainty
    (suppressed where it is negligible compared to the significant digits
    shown).
  }
  \label{tab:cross-sections}
\end{table}
\begin{figure}
  \centering
  \includegraphics[width=0.34\textwidth]{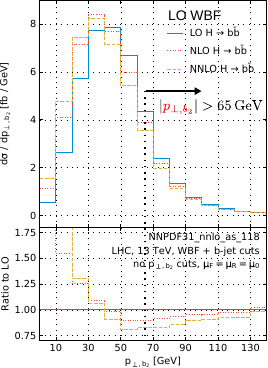}
  \qquad
  \includegraphics[width=0.34\textwidth]{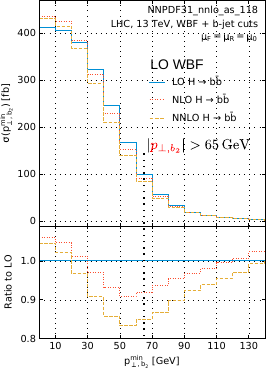}
  \caption{Fixed-order results for the transverse momentum distribution of the
    second hardest $b$ jet. All fiducial cuts apply except for the cut on
    $p_{\perp,b}$. The WBF production process is calculated at $\LO$, while the
    different lines show results for the decay process calculated at $\LO$,
    $\NLO$, and $\NNLO$ respectively. The left panel shows the transverse
    momentum distribution, while the right panel shows the cumulative
    distribution, i.e., the distribution integrated from
    $p_{\perp,b_2}^\text{min}$ to infinity.
    Plots based on Ref.~\cite{Asteriadis:2024nbg}.}
  \label{fig:fo-pTb2-decay}
\end{figure}
In Ref.~\cite{Asteriadis:2024nbg}, we combined the fixed-order calculations at
$\NNLO$ in QCD for WBF Higgs production from Ref.~\cite{Asteriadis:2021gpd} and
for $H \to b \bar{b}$ decays from Ref.~\cite{Behring:2019oci}.
We observe that there are surprisingly large corrections, both to the fiducial
cross section and to differential distributions, when taking into account
$\NNLO$ QCD corrections. The fiducial cross sections calculated with this setup
are listed in the first column of \cref{tab:cross-sections}.
Going from $\LO$ to $\NLO$, the central value of the fiducial cross section
decreases by about $\qty{31}{\percent}$, and when going from $\NLO$ to $\NNLO$
by another roughly $\qty{10}{\percent}$.
Similar observations hold for the corrections to differential distributions.
The size of the QCD corrections calls into question the reliability of the
perturbative expansion for observables using the fiducial cuts described above.

We were able to trace a significant portion of these effects to contributions
from higher-order corrections to the decay. As we already alluded to above, the
fiducial cuts on the $b$ jets originating from the Higgs boson decay are rather
strict. As a consequence, QCD radiation from the decay that carries away some
of the transverse momentum of the decay products can reduce the transverse
momentum of the $b$ jets to below the transverse momentum cut.
This is supported by the transverse momentum spectrum of the second-hardest $b$
jet, see \cref{fig:fo-pTb2-decay}, shown there \emph{without} the $p_{\perp,b}
> \qty{65}{\GeV}$ cut. The cut at $p_{\perp,b}^\text{cut} = \qty{65}{\GeV}$ sits
at a point where the spectrum is steeply falling, and the bulk of the spectrum
is located below this cut. Therefore, slight shifts to the position of the peak
or changes to the slope of the spectrum from radiative corrections can have a
large impact on fiducial observables.
This led to the hypothesis that a major part of these corrections can be taken
into account by adding a parton-shower simulation to the decay.

\section{Parton-shower-matched calculation}
In order to test this hypothesis, we replaced the fixed-order calculation for
the Higgs boson decay by a parton-shower-matched calculation based on
Ref.~\cite{Bizon:2019tfo}. This calculation uses \texttt{POWHEG}~\cite{%
Nason:2004rx,Frixione:2007vw,Alioli:2010xd} and the \texttt{MiNLO}
method~\cite{Hamilton:2012np,Hamilton:2012rf} to match the parton shower of
\texttt{Pythia 8.3}~\cite{Bierlich:2022pfr} to an $\NNLO$-accurate fixed-order
calculation.

\begin{figure}
  \centering
  \includegraphics[width=0.34\textwidth]{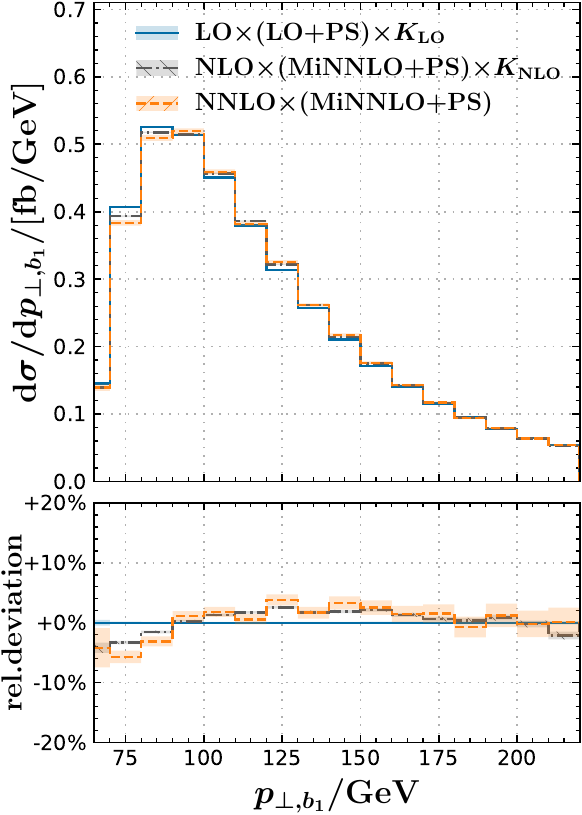}
  \qquad
  \includegraphics[width=0.34\textwidth]{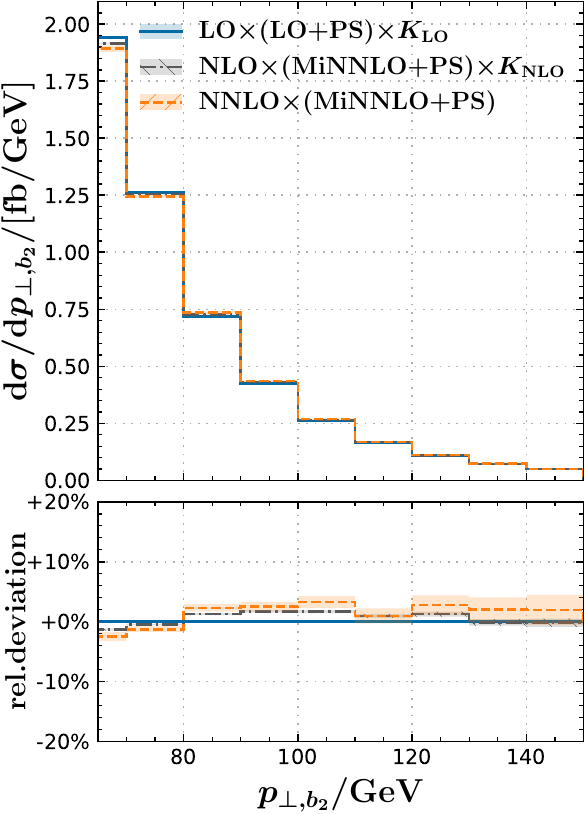}
  \caption{Results matched to parton shower for the transverse momentum 
    distributions of the hardest (left panel) and the second-hardest (right
    panel) $b$ jet. The lower-order distributions have been rescaled to the
    normalisation of the highest available order (see main text for details).
    The shaded error bands show only the uncertainties from numerical Monte
    Carlo integration.
    Plots from Ref.~\cite{Behring:2025msh}.}
  \label{fig:ps-ptb-withKfact}
\end{figure}
\begin{figure}
  \centering
  \includegraphics[width=0.34\textwidth]{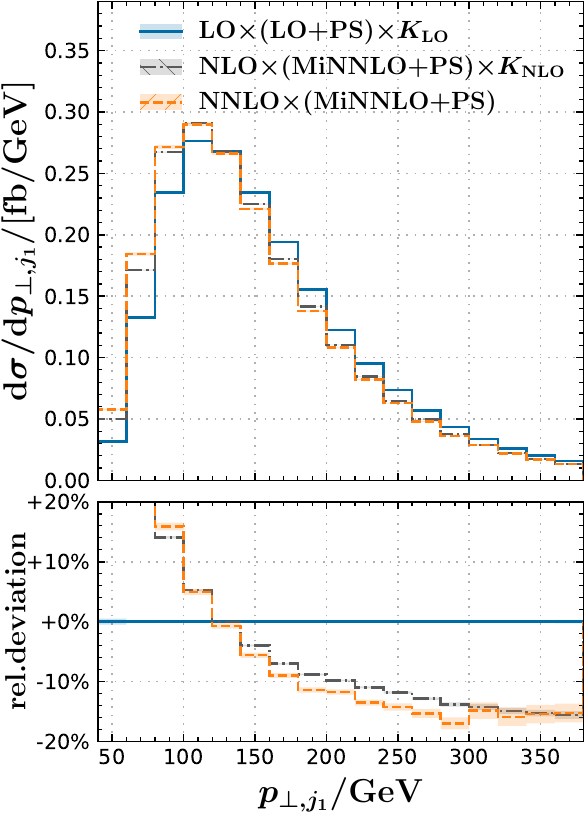}
  \qquad
  \includegraphics[width=0.34\textwidth]{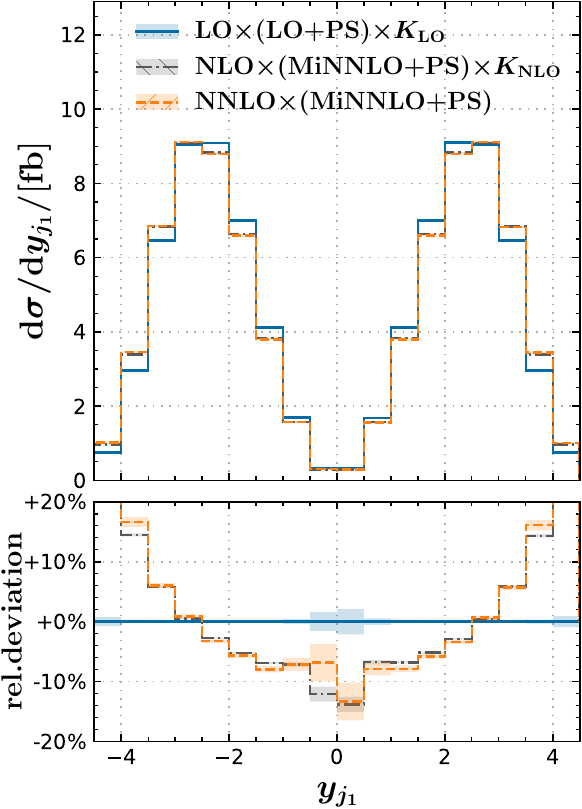}
  \caption{Results matched to parton shower for the transverse momentum and
    rapidity distributions of the hardest light jet. The lower-order
    distributions have been rescaled to the normalisation of the highest
    available order (see main text for details).
    The shaded error bands show only the uncertainties from numerical Monte
    Carlo integration.
    Plots from Ref.~\cite{Behring:2025msh}.}
  \label{fig:ps-wbfj-withKfact}
\end{figure}
The central values for the fiducial cross sections using an $\LOPS$ or an
$\MiNNLOPS$ approximation for the decay subprocess are shown in the second and
third columns of \cref{tab:cross-sections}, respectively. The fiducial cross
section at $\NNLO \times (\MiNNLOPS)$ is about $\qty{11}{\percent}$ smaller than
at $\LO \times (\LOPS)$.
Compared to the fixed-order calculation, where going from $\LO \times \LO$ to
$\NNLO \times \NNLO$ shifts fiducial cross sections by about
$\qty{41}{\percent}$, already the $\LO \times (\LOPS)$ results capture the
majority of the higher-order corrections. Going to the highest available order
for both the production and decay, we see that the $\NNLO$ results are fairly
stable under the inclusion of parton shower effects.

We now turn to differential distributions calculated using the
parton-shower-matched calculation.
\Cref{fig:ps-ptb-withKfact,fig:ps-wbfj-withKfact} show results obtained using
different approximations. The distributions at $\LO \times (\LOPS)$ and $\NLO
\times (\MiNNLOPS)$ have been rescaled using a fiducial $K$ factor defined via
$K_i = \frac{\sigma_{\NNLO \times (\MiNNLOPS)}}{\sigma_i}$ to normalise the
fiducial cross section to the best available approximation.

Similar to the fiducial cross sections, the differential distributions of
$b$-jet observables such as the transverse momenta of the leading and
subleading $b$ jet (see \cref{fig:ps-ptb-withKfact}) are already well described
by the $\LO \times (\LOPS)$ calculation. Going to $\NNLO \times (\MiNNLOPS)$
induces changes at the level of a few percent.
On the other hand, when considering observables that are primarily related to
the WBF production process, such as the transverse momentum of the hardest
WBF-tagging jet or its rapidity (see \cref{fig:ps-wbfj-withKfact}), substantial
higher-order corrections remain when going to $\NNLO$.
We find the dominant source of uncertainty to be the renormalisation scale
variation in the decay subprocess, which in the $\MiNNLOPS$ calculation amounts
to about $\qty{5}{\percent}$ in most regions of phase
space~\cite{Behring:2025msh}. Renormalisation and factorisation scale variation
in the production subprocess are of the order of $\qty{2}{\percent}$~\cite{%
Asteriadis:2024nbg}, leading to an overall uncertainty estimate of around
$\qtyrange{5}{7}{\percent}$ for our $\NNLO \times (\MiNNLOPS)$ calculation.

\section{Conclusions}
With the two studies described here, we investigated the effect of higher-order
corrections to Higgs production in the WBF production channel with subsequent
decays of the Higgs to bottom-quark pairs under realistic fiducial cuts. We
observe large perturbative corrections of up to about $40\%$ relative to
leading order in an $\NNLO$ QCD fixed-order calculation.
We trace back these corrections to an interplay between radiative corrections
to the Higgs decay and the transverse momentum cut on $b$ jets.
Including a parton shower description for the Higgs decay, matched to an
$\NNLO$ QCD calculation, effectively resums these large corrections and
improves the perturbative convergence.
The predictions could be further improved, e.g., by including electroweak
corrections and by extending the parton shower description to both the
production and decay processes. Nonetheless, our studies demonstrate the
intricacies of precise predictions for observables in fiducial phase spaces and
the merit of including parton shower matching to capture effects of radiative
corrections.

\bibliographystyle{JHEP}
\bibliography{proceedings}

\end{document}